\documentclass[a4paper]{jpconf}
\usepackage{graphicx}
\begin{document}

\title{Dependence of vortex phase transitions in mesoscopic
Bi$_{2}$Sr$_{2}$CaCuO$_{8}$ superconductor at tilted magnetic
fields}

\author{$^1$ M I Dolz and $^2$ H Pastoriza}

\address{Centro At\'omico Bariloche, Comisi\'on Nacional de Energía At\'omica and CONICET, R8402AGP S.
C. de Bariloche, Argentina}

\ead{$^1$ mdolz@cab.cnea.gov.ar, $^2$ hernan@cab.cnea.gov.ar}

\begin{abstract}
A micron sized single crystal of the superconductor
Bi$_{2}$Sr$_{2}$CaCuO$_{8}$ was studied using silicon mechanical
micro-oscillators at various tilt angles of the \emph{dc} magnetic
field with respect to the \emph{c} axis of the sample. Different
phases of the vortex matter were detected by measuring changes in
the value and sign of the oscillator resonant frequency variation
with temperature. We could explain the change in the sign of this
variation at high temperatures as the transition from the $2D$
liquid of decoupled pancakes to a reversible $3D$ vortex lattice.
The data indicates that this transition only depends on the
magnetic field perpendicular to the superconducting layers while
the dissipation involved in this process depends on the component
parallel to them.
\end{abstract}

\section{Introduction}

The vortex matter structure in high $T_{c}$ superconductors is
determined by the competition between three important energy
scales: thermal, repulsive interaction between vortices and
attractive interactions between them and materials defects
(pinning). The Bi$_{2}$Sr$_{2}$CaCuO$_{8}$ (BSCCO) is a strongly
anisotropic type-II superconductor and because of all these
factors it presents an ample, complex and rich phase diagram with
numerous phase transitions and crossovers. A magnetic field, $H$,
perpendicular to the superconducting planes ($H_{c}$ parallel to
$c$ axis) penetrates the sample as $2D$ pancakes vortices, PV`s,
\cite{clem} while a transverse field ($H_{ab}$ parallel to $ab$
planes) penetrates as Josephson vortices, JV`s. The JV`s couple
PV`s from different superconducting planes and this connection
disappears at high temperatures where the lattice of PV`s melts
into a liquid. Under tilted fields this transition in macroscopic
samples has been studied intensively \cite{tilted}. In this paper
we show that in a micron sized sample of BSCCO this transition
only depends on the magnetic field perpendicular to the
superconducting layers ($H_{c}$) although the dissipation involved
in this process depends on the component parallel to them
($H_{ab}$).

In order to detect the small signals produced by microsized
samples, new instruments designed for the microscale are needed.
Our approach for studying the magnetic properties of a
Bi$_2$Sr$_2$CaCu$_2$O$_{8 + \delta}$ microscopic disk is to use
silicon micro-oscillators (following the work of \cite{bolle99}
and \cite{dolz07}) which have a torsional mode with a resonant
frequency $\nu _r \approx 45$ kHz and a quality factor $Q > 10^4$
at low temperatures. This instrument integrates high sensitivity
and reduced size with a small signal loss.

\section{Experimental}

The poly-silicon oscillators were fabricated in the MEMSCAP
\cite{memscap} foundry using its Multiuser process (MUMPS). The
oscillator consists of a 50 $\times$ 100 $\mu$m$^{2}$ released
plate anchored to the substrate by two serpentine springs. A more
detailed description of the experimental detection setup can be
found elsewhere \cite{dolz07,dolz08}. Single crystals of BSCCO
were grown using the self flux technic \cite{gladys} and by means
of optical lithography and ion milling we fabricated disks of 13.5
microns in diameter and 1 micron in high. A disk was mounted on
the micro-oscillator`s plate with the Cu-O layers ($\emph{ab}$
planes) parallel to it (figure 1a). The measurements were taken in
vacuum inside a closed-cycle cryogenerator where the temperature
can be varied between $14$ and $300\,$K. The $dc$ magnetic field
was provided by a split electromagnet that can be rotated in the
plane perpendicular to the axis of rotation of the oscillators
with an accuracy of 1$\,^{\circ}$. Therefore, the direction of $H$
can be varied from the $\emph{c}$ axis of the sample ($H=H_{c}$)
to an axis perpendicular to it ($H=H_{ab}$).

\vspace{1cm}

\begin{figure}[h]
\begin{minipage}{7.5cm}
\begin{center}
\includegraphics[width=6.5cm]{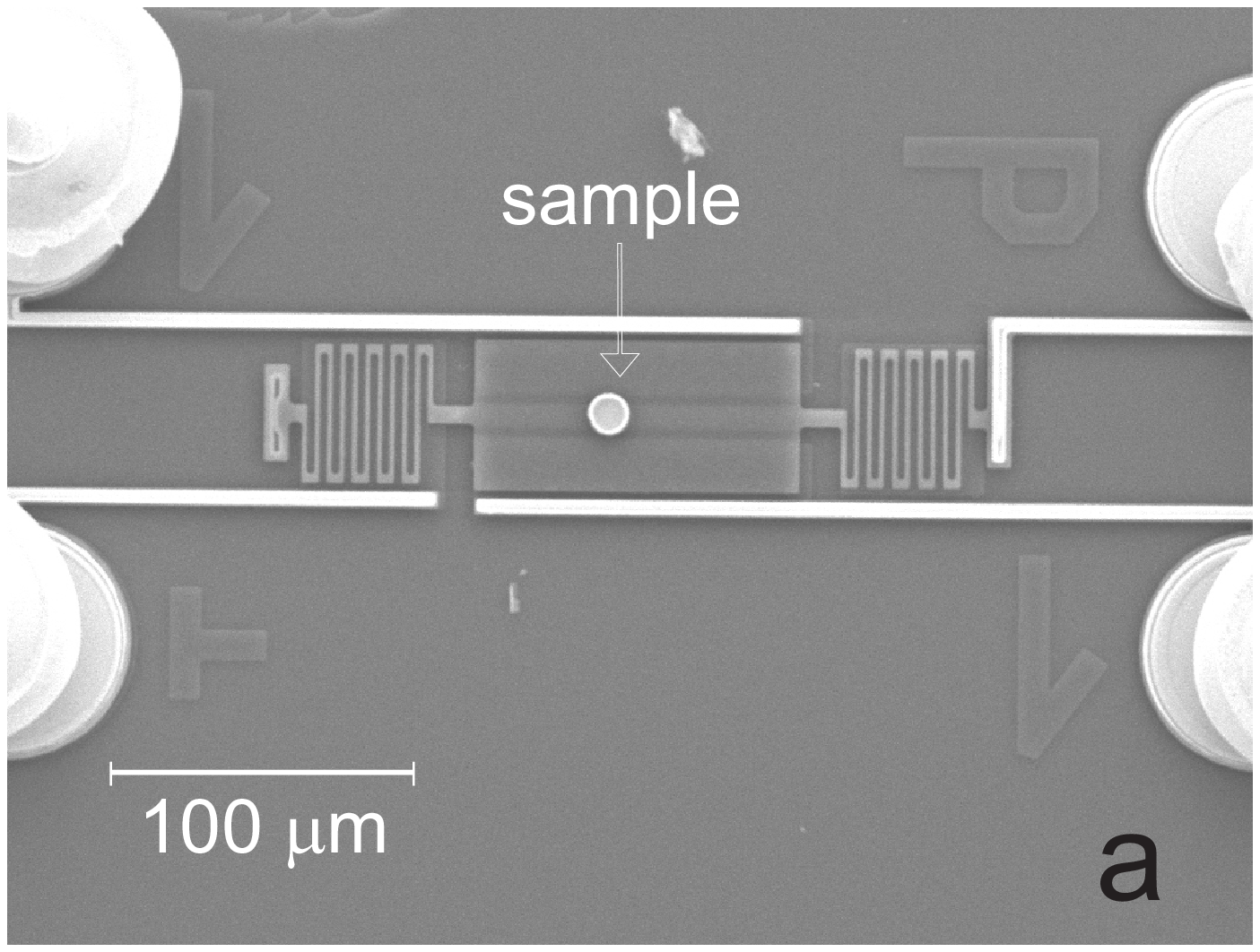}
\end{center}
\end{minipage}\hspace{0.3cm}%
\begin{minipage}{7.5cm}
\begin{center}
\includegraphics[width=7.5cm]{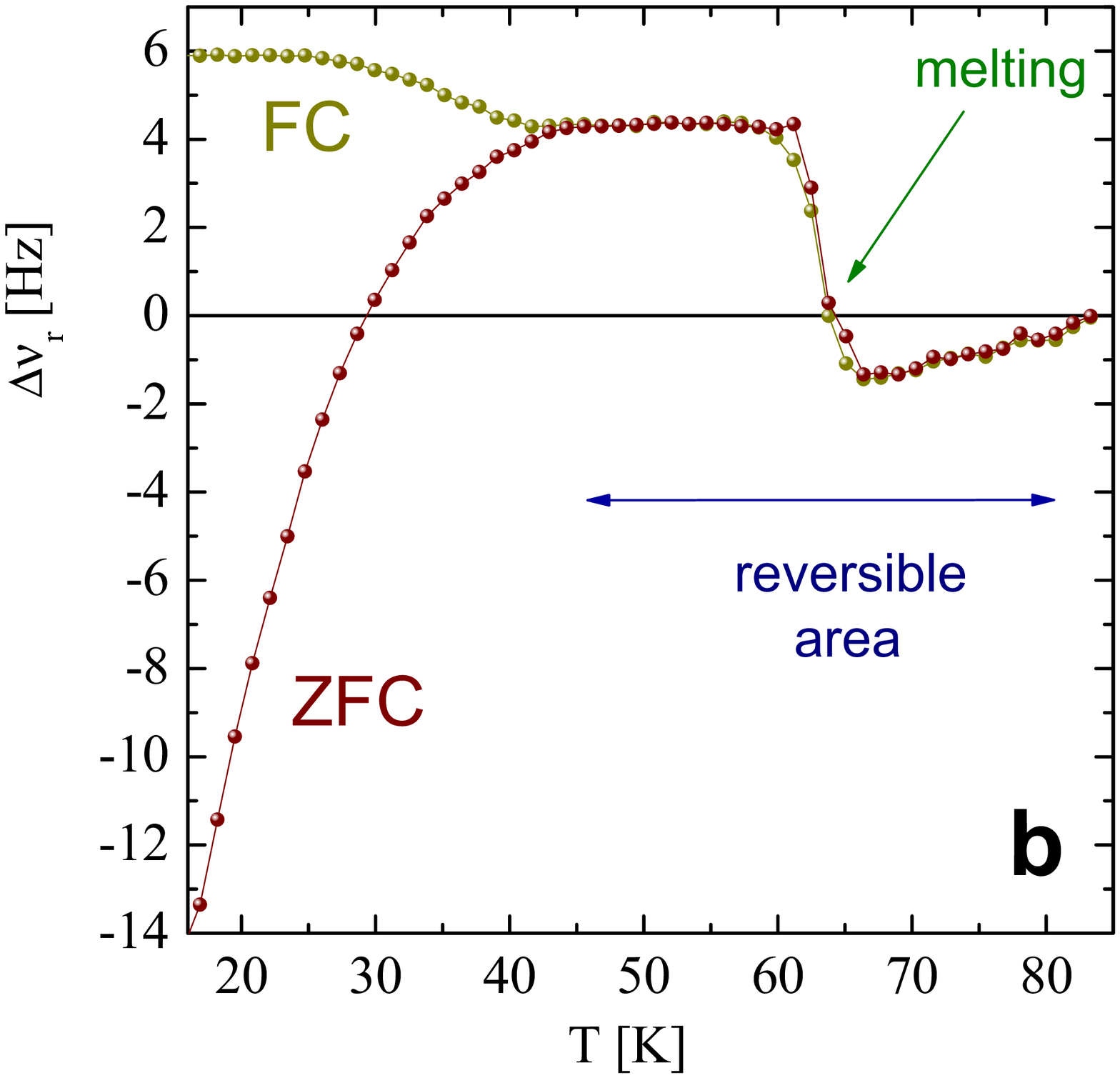}
\end{center}
\end{minipage}
\caption{a) Scanning electron micrograph of a high-Q mechanical
silicon oscillator with a BSCCO disk with a diameter of $13,5 \mu
m$ and a thickness of $1 \mu m$. b) Change in the resonant
frequency as a function of temperature under a tilted external
magnetic field.}
\end{figure}
\vspace{1cm}

The natural resonant frequency ($\nu_0$) of an oscillator in the
torsional mode is given by: $ 2\pi\nu_0=\sqrt{\frac{k_e}{I}},$
where $k_e$ is the elastic restorative constant of the serpentine
springs and $I$ is the plate moment of inertia. In our system
(oscillator+sample) this mode has a resonant frequency close to
$40$ kHz and a quality factor $Q$ greater than $2.6\times 10^{4}$.
When a magnetic sample is attached to the oscillator and $H$ is
applied the resonant frequency $\nu_r$ changes to: $2\pi \nu_{r} =
\sqrt{\frac{k_e+k_M}{I}}$ where $k_M$ is the variation in the
effective elastic constant originated by the magnetic interaction
between the sample and the magnetic field. It can be expressed as:
\begin{equation}
k_M \simeq 8 \pi^2 I \nu_0 \Delta\nu,
\end{equation}
where $\Delta\nu=\nu_0-\nu_r$ is the change in the resonant
frequency when $H$ is applied. This $k_M$ is associated with the
sample magnetization, $M$, and can be calculated by
differentiating two times the magnetic energy, as in
\cite{dolz08}. The experiment consist in measuring the
oscillator`s torsional mode resonant frequency as a function of
temperature to different values and directions of $H$.

\section{Results and Discussions}

In the two limit cases where $H=H_{c}$ and $H=H_{ab}$ we showed
\cite{dolz07} that $\Delta \nu_r$ is always negative and positive,
respectively. In the first case this is due to the non-restorative
torque between $H_{c}$ and the magnetic moment generated by
Meissner currents. In the second case the restorative torque is
generated by the interaction between $H_{ab}$ and the
superconducting currents that shield the alternating perpendicular
magnetic field $h_{c}=H_{ab}\times\sin\alpha$, where $\alpha$ is
the angle of the oscillator with respect to its equilibrium
position. On the other hand, the response of the system to tilted
fields $H=H_{\theta}$, where $\theta$ is the angle between the
field and $\emph{c}$ axis of the sample, can be positive or
negative depending on the vortex system phase. Figure 1b shows ZFC
(zero field cooling) and FC (field cooling) measurements for $H$
at $\theta=30\,^{\circ}$. The presence of $ac$ magnetic fields
generated by the tilt of the micro-oscillator, causes that $\Delta
\nu_r$ is related with the $dc$ magnetization, $M$, and the $ac$
response, $\chi$, of the sample. It is known that the melting
transition imply features in $M$ and $\chi$ \cite{doyle95}. The
only peculiar or characteristic behavior in our measurements at
high temperatures is the change of sign of $\Delta \nu_r$, and in
this way we can relate it with the melting transition from $2D$
liquid of decoupled pancakes to a reversible $3D$ vortex lattice.
On the other hand, from low temperatures at the irreversible area
the non-restorative torque in ZFC measurements is basically due to
the Meissner currents and for FC measurements the response $\Delta
\nu_r > 0$ is due to the pinning`s vortices.

\vspace{1cm}

\begin{figure}[h]
\begin{minipage}{7.5cm}
\begin{center}
\includegraphics[width=7.5cm]{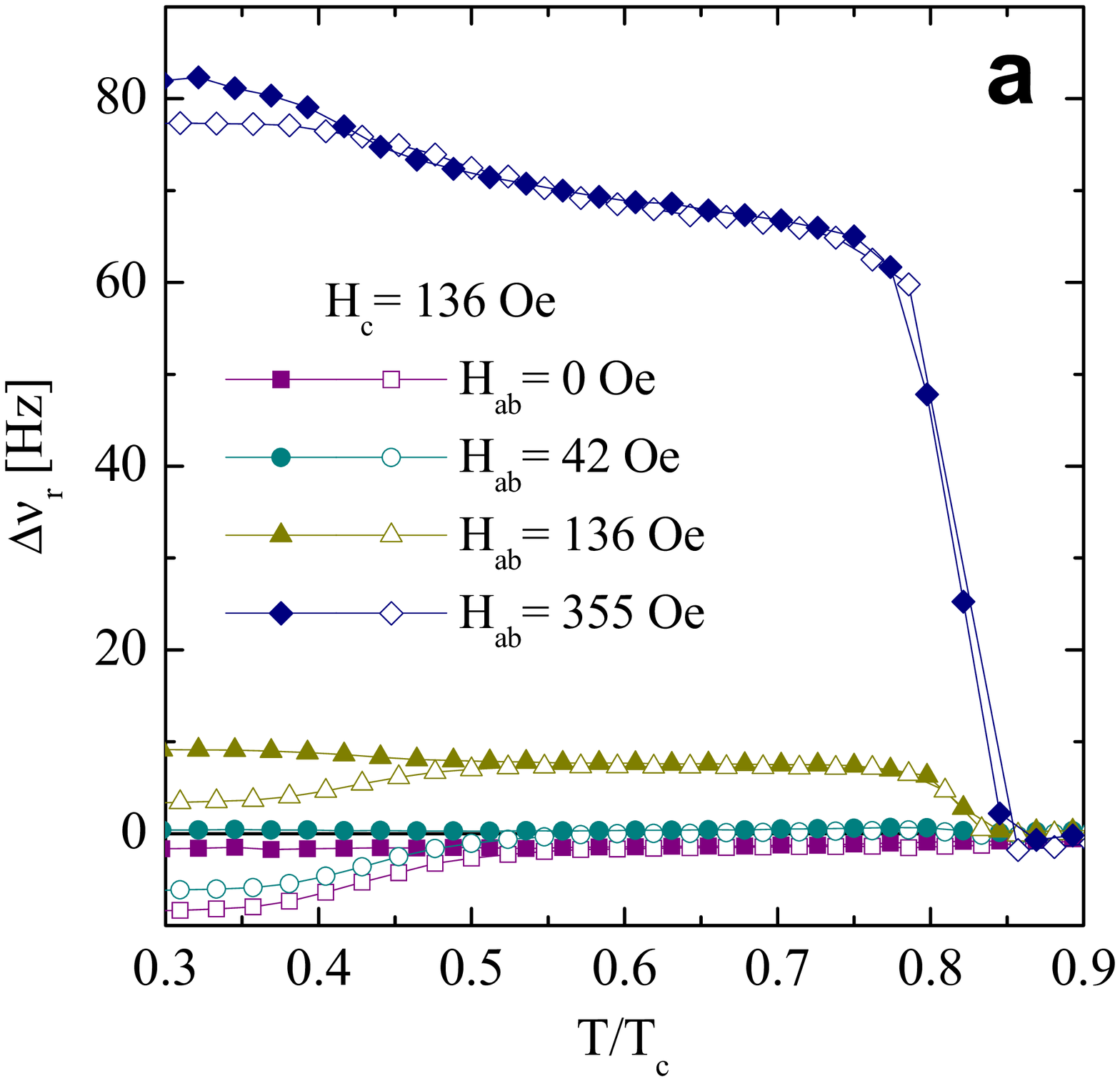}
\end{center}
\end{minipage}\hspace{0.3cm}%
\begin{minipage}{7.5cm}
\begin{center}
\includegraphics[width=7.5cm]{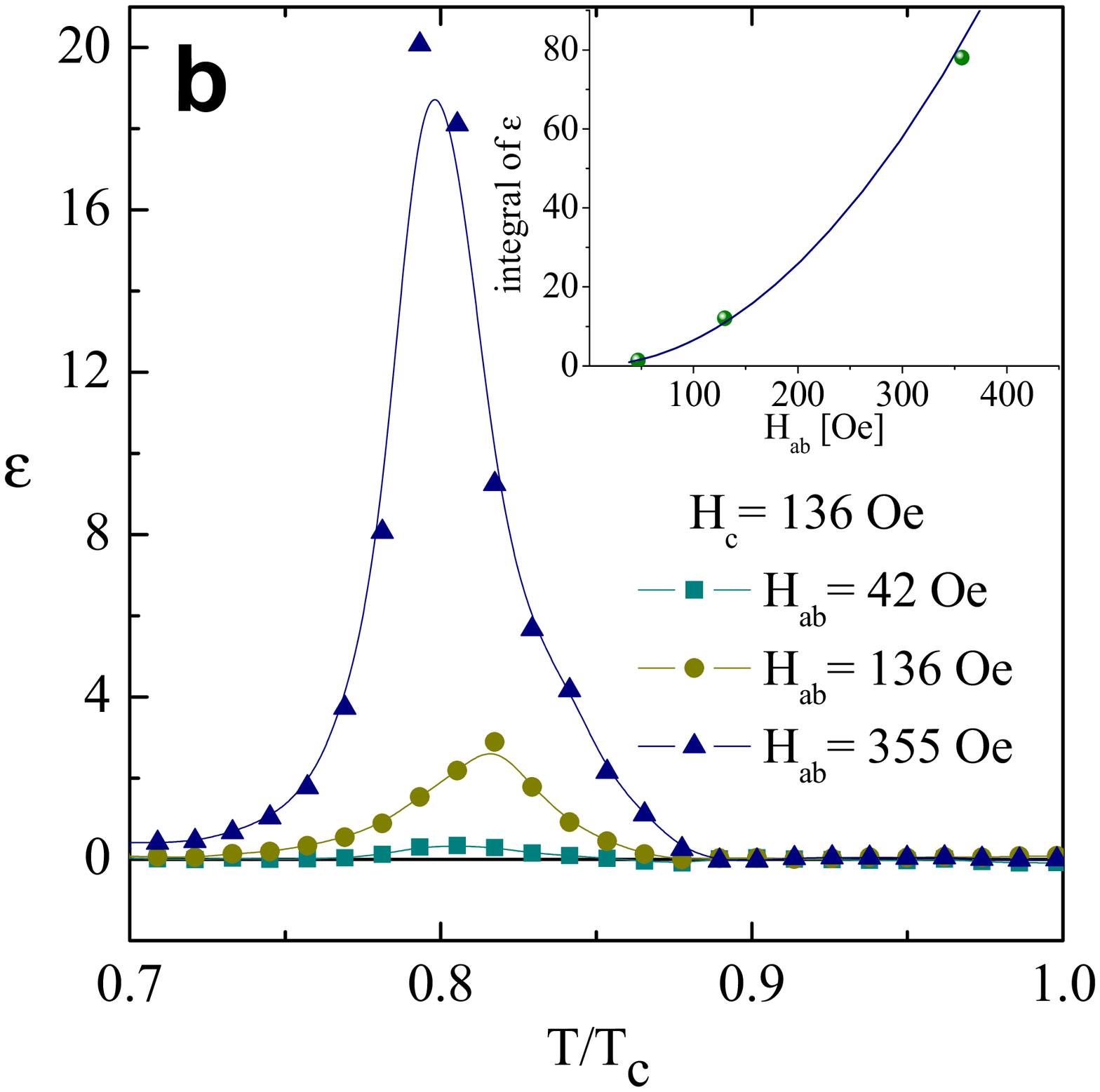}
\end{center}
\end{minipage}
\caption{a) $\Delta \nu_r$ vs reduced temperature applying a
constant magnetic field perpendicular to the superconducting
planes for different values of $H_{ab}$ as indicated in the
figure. b) Dissipated energy of the sample vs reduced temperature.
Inset: Integral of the dissipation at the melting transition vs
$H_{ab}$.}
\end{figure}

\vspace{1cm}

The measured $\Delta \nu_r$ vs. $T$ data for different angles and
values of $H$, keeping constant $H_{c}=136\ \textrm{Oe}$ is showed
in figure 2a. We observe that for microscopic BSCCO samples the
melting transition occurs at the same reduced temperature $t \sim
0.815$ and does not depends on the applied $H_{ab}$. This result
agrees with the conclusions obtained in ref \cite{kes90} who
proposed that in BSCCO, the phase transition from the vortex
liquid phase to the vortex solid state depends only on $H_{c}$.

Due the Lorentzian fit of the amplitude vs. frequency curve from
which we obtain $\nu_r$, we also can obtain $Q$. In this way we
can find the energy, $\epsilon$, dissipated for the sample by
cycle in each temperature (figure 2b):

\begin{equation}
\epsilon \propto Q_{0}/Q_{H}-1,
\end{equation}

where $Q_{H}$ and $Q_{0}$ are the quality factors with and without
$H$. The integral of $\epsilon$ over the range of temperature
where the transition is present, is a quadratic function of
$H_{ab}$ (inset figure 2b). While the melting temperature no
depends on $H_{ab}$ the jump in $\Delta \nu_r$ does it. The term
of $\Delta \nu_r$ (or of energy) related with the $ac$ currents
shielding $h_{c}$ is a quadratic function of $H_{ab}$. This means
that the dissipation of the system is controlled only by $ac$
currents generate in the superconducting planes.

\section{Conclusions}

In conclusion, we have presented measurements of a micron sized
single crystal of the superconductor Bi$_{2}$Sr$_{2}$CaCuO$_{8}$
using silicon mechanical micro-oscillators. The change in the
resonant frequency can be positive or negative, depending on how
the currents and magnetic moments are generated in the sample. The
melting transition can be detected as the change of sign in
$\Delta \nu_r$ at high temperatures and it is independent of the
value of the magnetic field parallel to the superconducting planes
and only depends on the magnetic field perpendicular to them, in
agreement with \cite{kes90}. However, at the transition the
dissipated energy is a quadratic function of $H_{ab}$, indicating
that the $ac$ currents in the superconducting planes are the most
important source of dissipation.


\section*{References}
\begin{thebibliography}{9}
\bibitem{clem} Clem J R 1991 \emph{Phys. Rev. B} \textbf{43} 7837
\bibitem{tilted} Mirkovic J, Savelev S E, Sugahara E and Kadowaki K 2001 \emph{Phys.
Rev. Lett.} \textbf{86 } 886. Mirkovic J, Savelev S E, Sugahara E
and Kadowaki K 2001 \emph{Physica C} \textbf{357–360} 450.
Grigorenko A et al. 2001 \emph{Nature} \textbf{414} 728. Koshelev
A E 1999 \emph{Phys. Rev. Lett.} \textbf{83} 187. Savelev S E,
Mirkovic J and Kadowaki K 2001 \emph{Phys. Rev. B}\textbf{64}
94521
\bibitem{bolle99} Bolle C A, Aksyuk V, Pardo F, Gammel P L, Zeldov E, Bucher E, Boie R, Bishop D J and Nelson D R 1999 \emph{Nature}
\textbf{399} 43
\bibitem{dolz07} Dolz M I, Antonio D and Pastoriza H 2007 \emph{Physica B} \textbf{398} 329–332
\bibitem{memscap} MEMSCAP Inc., Durham, NC, USA. http://www.memscap.com
\bibitem{dolz08} Dolz M I, Bast W, Antonio D, Pastoriza H, Curiale J, Sánchez R D and Leyva A G 2008 \emph{J. Appl. Phys. } \textbf{103}
083909
\bibitem{gladys} Kaul E and Nieva G 2000 \emph{Physica C} \textbf{341-348} 1343-1344
\bibitem{doyle95} Doyle R A, Liney D, Seow W S and Campbell A M
1995 \emph{Phys. Rev. Lett.} \textbf{75} 4520.
\bibitem{kes90} Kes P H, Aarts J, Vinokur V M and van
der Beek C J 1990 \emph{Phys. Rev. Lett}. \textbf{64} 1063


\end{thebibliography}
\end{document}